\documentclass[journal]{IEEEtran}

\RequirePackage{fix-cm}

\usepackage{graphicx}
\usepackage{subfig}

\usepackage[hidelinks]{hyperref}
\usepackage{xcolor}  
\usepackage{amsmath,amssymb,amsthm,mathtools}

\DeclareMathOperator*{\argmin}{arg\,min}

\newcommand{\sumt}{\sum_{t\in \mathcal{T}}}

\newcommand{\sumder}{\sum_{j \in \mathcal{N}_{\text{DER}}}}
\newcommand{\fat}{\forall t \in \mathcal{T}}
\newcommand{\fap}{\forall \phi \in \Phi_j}

\newcommand{\R}{\mathcal{R}}
\newcommand{\T}{\intercal}
\renewcommand{\b}{\boldsymbol}
\usepackage{framed}
\usepackage{etoolbox}
\let\bbordermatrix\bordermatrix
\patchcmd{\bbordermatrix}{8.75}{4.75}{}{}
\patchcmd{\bbordermatrix}{\left(}{\left[}{}{}
\patchcmd{\bbordermatrix}{\right)}{\right]}{}{}

\usepackage[ruled,vlined]{algorithm2e}

\usepackage[font=footnotesize]{caption}

\theoremstyle{definition}

\makeatletter
\newcommand{\nosemic}{\renewcommand{\@endalgocfline}{\relax}}
\newcommand{\dosemic}{\renewcommand{\@endalgocfline}{\algocf@endline}}
\makeatother
\usepackage{framed}
\usepackage{arydshln}
\begin{document}
\graphicspath{{./artwork/}}  

\title{Valuing Distributed Energy Resources for Non-Wires Alternatives}
%
%
%

\author{Nicholas~D.~Laws,~\IEEEmembership{Student~Member,~IEEE,}
        and~Michael~E.~Webber
\thanks{N. D. Laws is with the Walker Department of Mechanical Engineering, The University of Texas at Austin, TX, USA (e-mail: nlaws@utexas.edu).}
\thanks{M. E. Webber is with the Walker Department of Mechanical Engineering, The University of Texas at Austin, TX, USA  (e-mail: webber@mail.utexas.edu).}
\thanks{Manuscript received December 14, 2021.}}

%
%

\markboth{IEEE Transactions on Smart Grid,~Vol.~X, No.~X, December~2022}%
{Laws \MakeLowercase{\textit{et al.}}: Valuing Distributed Energy Resources for Non-Wires Alternatives}
\maketitle

\begin{abstract}
  Distributed energy resources (DER) as non-wires alternatives, regardless of owner, have the potential to reduce system operating costs and delay system upgrades. 
  However, it is difficult to determine the appropriate economic signal to incentivize DER investors to install capacity that will benefit both the DER investors and the system operator. 
  In an attempt to determine this co-optimal price signal, we present a bilevel optimization framework for determining the least cost solution to distribution system over-loads. 
  A key output of the framework is a spatiotemporal price signal to DER owners that simultaneously guarantees the DER owners' required rate of return and minimizes the system operation costs. 
  The framework is demonstrated with a case by which the system operator considers utility owned battery energy storage systems, traditional system upgrades, and energy purchase from DER owners. The results show that by valuing DER for non-wires alternatives the utility owned storage system sizes can be reduced, less hardware upgrades are necessary, and upfront capital costs as well as operating costs are reduced.
\end{abstract}


%
\IEEEpeerreviewmaketitle

  \begingroup
  \begin{table}[th!] 
  \caption{Nomenclature}
  \label{tab:args}
  \renewcommand{\arraystretch}{1.3} 
  \begin{tabular}{p{0.6in}p{2.5in}}
    \multicolumn{2}{l}{\textbf{Decision Variables}} 
    \\ \hline \hline
      $\b{x} \in \R^M$ & upper level primal decision variables
    \\
      $\b{y} \in \R^N$ & lower level primal decision variables
    \\
      $\b{z} \in \lbrace 0,1 \rbrace^K$ & upper level primal binary decision variables
    \\
      $\b{\lambda} \in \R^J$ & lower level, dual variables for equality constraints
    \\
      $\overline{\b{\mu}} \in \R^N_+$ & lower level dual variables for upper bounds
    \\
    $\underline{\b{\mu}} \in \R^N_+$ & lower level dual variables for lower bounds
  \\
    \multicolumn{2}{l}{\textbf{Parameters}} 
    \\ \hline \hline
      $ \b{V} \in \R^{J \times N}$  & lower level equality constraint coefficients
    \\
      $\b{w} \in \R^J$  & lower level equality constraints right-hand-side
    \\
    $\overline{\b{y}} \in \R^N$ & upper bounds for lower level, primal decision variables
    \\
    $\underline{\b{y}} \in \R^N$ & lower bounds for lower level, primal decision variables
    \\
    $a$  & upper level scaling coefficient for cost of DER energy
    \\
      $b$  & lower level scaling coefficient for income from selling energy to upper level
    \\
      $\b{c} \in \R^N$  & lower level cost coefficients for lower level decisions $\b{y}$
    \\
    \multicolumn{2}{l}{\textbf{Sets and Indices}} 
    \\ \hline \hline
      $\mathcal{E}$  & set of edges in the network
    \\
      $\mathcal{N}$  & set of nodes in the network
    \\
      $\mathcal{N}_{\text{DER}}$  & set of nodes for potential DER investors
    \\
      $\mathcal{N}_{\text{BESS}}$  & set of nodes for potential BESS installations
    \\
      $\mathcal{N}_{\text{TRFX}}$  & set of nodes for potential transformer upgrades
    \\
      $\mathcal{N}_{\text{LINE}}$  & set of nodes for potential line upgrades
    \\
      $\mathcal{S}$  & set of demand charge periods
    \\
      $\mathcal{T}$  & set of time steps
    \\
      $\Phi_j$  & set of phases connected to node $j$
  \end{tabular}
  \end{table}
  \endgroup

\section{Introduction and Background}\label{sec:background}
As population and electrification grow, there is pressure on local electric distribution utilities to increase overall capacity and upgrade equipment to maintain high reliability. These actions include simple maintenance such as vegetation management, but also replacing and upgrading transformers, installing more lines, replacing lines with newer ones, and so forth. 

However, those traditional actions related to the wires and poles of the distribution system might not keep pace with load growth that will accommodate rapid electric vehicle adoption or widespread installation of electric heat pumps as a way to reduce on-site fuel use for space and water heating. As a consequence, there is an acute need for non-wires alternatives that can be used to improve overall system performance. Some of those alternatives include demand response and distributed energy resources, such as local power generation and/or storage.

Though distributed energy resources avoid additional loading on the distribution lines, traditional utility funding models do not always support their installation. Furthermore, market signals can be confusing and do not encourage DER installation even though novel business models are emerging that would reduce total system cost.

Given this context, this research seeks to explain how DER might be appropriately valued in light of increasing electrification and EV adoption in an era when traditional grid enhancements are hobbled by cost and policy hurdles.

Early evaluations of DER for non-wires alternatives compared costs and benefits of known DER capacities and locations against capacity upgrade costs \cite{piccolo2009evaluating}. A common theme in the literature for valuing DER as non-wires alternatives accounted for the single perspective of the distribution system operator (DSO).
For example, Contreras-Ocana \textit{et al}. developed a model that puts DER costs and benefits in competition with upgrade deferrals from a single perspective, at a single location (substation or transformer) with forecasted overloads \cite{contreras2019non}. By neglecting power flow constraints they were able to account for many types of DER including energy efficiency investments. However, without a network model the DER are presumably installed at the single, overloaded location.

The valuable work by Andrianesis \textit{et al}. demonstrates how to determine a locational marginal value (LMV) of DER in a three step process, where the value of DER is determined relative to the locational marginal cost of traditional system upgrades \cite{andrianesis2019locational}. 
The method in \cite{andrianesis2019locational} also only accounts for the DSO perspective, implicitly assuming that DER will behave in a manner that best suits the system operator's cost function and constraints. Furthermore, it is assumed that the LMV is sufficient to motivate DER investment and that the LMV will not motivate behavior that leads to greater operating costs or the need for additional system upgrades.


The work by Garcia-Santacruz \textit{et al}. is perhaps the first to account for both the DSO and DER owner perspectives when valuing DER for non-wires alternatives \cite{garcia2021}. 
The DSO perspective is represented by minimizing energy losses in a relaxed branch flow model \cite{farivar2013branch}. The DER owner perspective is represented by maximizing a DER utilization factor, i.e. minimizing the amount of energy curtailed. 
The system losses and DER utilization factors are combined in a single cost function with weighting factors that must be selected by the planner. 
To make their problem tractable, only 24 hours of loading is modeled, which prevents representation of the full value proposition of DER.

The method presented in this paper addresses several gaps in the literature for valuing DER as non-wires alternatives. 
First, by leveraging a bilevel optimization framework, we account for the perspectives of DER investors and the DSO without the need for ad-hoc weighting factors.
Second, we determine a co-optimal, spatiotemporal price signal that simultaneously minimizes DSO costs while guaranteeing the DER owners' financial returns.
And third, the optimal locations and capacities of both utility-owned and third-party-owned DER are determined while obeying power flow constraints.


The following sections present the proposed framework for valuing DER for non-wires alternatives and example applications of the framework. The examples use a year of hourly loads, which accounts for the full value proposition of DER for non-wires alternatives.

\section{Methodology}\label{sec:method}
To account for the objectives and constraints of \textit{both} the DSO and the potential DER investors we employ a bilevel optimization framework. Bilevel optimization problems, also known as Stackelberg Games, are generally intractable. However, by ensuring that the lower level problem is linear the bilevel problem can be converted into an equivalent single level problem \cite{dempe2020bilevel}. Furthermore, under certain conditions the bilinear products of dual variables, i.e. shadow prices, and primal variables can be linearized \cite{laws2022linearizing}. This last point is especially important since we seek to optimize the product of the price signal from the DSO and the power injection decisions of the DER investors.

The general bilevel optimization framework is shown in Problem \ref{equ:bilevel-general-linear} (with nomenclature in Table \ref{tab:args}).
\begin{subequations}  \label{equ:bilevel-general-linear}
\begin{align}
  \min_{\b x , \b y }  & \ 
  f(\b x, \b y)
    + \frac{a}{b} \sumder \sumt \lambda_{j,t} \ y^{\text{EXP}}_{j,t}  
    \label{equ:1a}
    \\
  \text{s.t.} & \  
      g(\b x, \b y) \le \b 0
    \\
      & \ \b{y} \in \argmin_{\b{y} \in \R^N}  \ \b{c}^\T \b{y} 
      - b \sumder \sumt x_{j,t}^{\lambda} \ y^{\text{EXP}}_{j,t}   \label{equ:gen-lower-obj}
    \\
      & \quad \quad\ \text{s.t.} \ \  \underline{\b{y}} \le \b{y} \le \overline{\b{y}} \quad \ (\underline{\b{\mu}}, \overline{\b{\mu}}) \label{equ:LL-bounds}
    \\
      & \qquad \quad \quad \b{V} \b{y} = \b{w} \quad (\b{\lambda}). \label{equ:LL-equality-constraints}
\end{align}
\end{subequations}
The components of the framework in (\ref{equ:bilevel-general-linear}) are:
\begin{itemize}
  \item the upper level cost (\ref{equ:1a}), which includes a generic function $f(\b x, \b y)$ and a second term that represents the payment to DER owners via the product of the lower level shadow price $\lambda_{j,t}$ and export decisions $y^{\text{EXP}}_{j,t}$;
  \item the upper level generic constraint set $g(\b x, \b y) \le \b 0$;
  \item the upper level is constrained by the lower level decisions $\b y$ being in the optimal space of the lower level problem;
  \item the lower level problem has a linear cost function given $\b x$ (\ref{equ:gen-lower-obj}), which includes:
  \begin{itemize}
    \item a generic cost $\b{c}^\T \b{y} $ and 
    \item the income from the upper level via price signal $x_{j,t}^{\lambda}$ times the exported DER energy $y^{\text{EXP}}_{j,t}$;
  \end{itemize}
  \item and the linear lower level constraints with the associated dual variables in parentheses (\ref{equ:LL-bounds})-(\ref{equ:LL-equality-constraints}).
\end{itemize}
The key feature of the framework is the exchange of money between the DSO in the upper level and the DER investors in the lower level via the products of the upper level price signal decisions and the lower level export decions in the objective functions (\ref{equ:1a}) and (\ref{equ:gen-lower-obj}). 
As we show in the following, the DSO in the upper level can effectively set the  marginal price of DER investors when the DSO wishes; and the sum of the total compensation paid to DER investors must meet the investors' required rate of return.

As shown in \cite{laws2022linearizing}\footnote{See Equation (9) in \cite{laws2022linearizing} when the index $k$ is equal to the index for a given $y^{\text{EXP}}_{j,t}$, the variable has zero coefficients in the cost vector $\b c$,  and no bounds are binding (i.e. $\underline{\b{\mu}} = 0$ and  $\overline{\b{\mu}} = 0$).}:
\begin{equation}
  \lambda_{j,t} = b \ x_{j,t}^{\lambda}
\end{equation}
which means that the upper level payment to DER can be written:
\begin{equation} \label{equ:3}
  \frac{a}{b} \sumder \sumt \lambda_{j,t} \ y^{\text{EXP}}_{j,t} 
  = a \sumder \sumt x_{j,t}^{\lambda} \ y^{\text{EXP}}_{j,t}.
\end{equation}
In other words, we have formed a zero-sum game in which the upper level, or DSO's, cost of DER energy is equal to the DER owners' income (with the exception of the scaling coefficients $a$ and $b$ that we will use to account for each party's cost of capital).
Furthermore, the product of the dual variable $\lambda_{j,t}$ and the DER dispatch variable $y^{\text{EXP}}_{j,t}$ can be shown to equal a linear sum \cite{laws2022linearizing}:
\begin{equation} \label{equ:linerization}
  \sumder \sumt 
  \lambda_{j,t} y^{\text{EXP}}_{j,t} 
  =
  \frac{1}{V_{y^{\text{EXP}}}} \left(
    \b w^\T \b \lambda
  - \b c^\T \b y 
  - \boldsymbol{ \overline{\mu}}^\T \boldsymbol{ \overline{y} }
  + \boldsymbol{ \underline{\mu}}^\T \boldsymbol{ \underline{y} }
  \right)
\end{equation}
where $V_{y^{\text{EXP}}}$ is the coefficient of the lower level export variable in the lower level load balance equations.

To convert the bilevel problem (\ref{equ:bilevel-general-linear}) into a single level problem it is important to note that the the lower level problem described by (\ref{equ:gen-lower-obj}) -- (\ref{equ:LL-equality-constraints}) is linear given $\b x$. 
The conversion to a single level is achieved by replacing the lower level problem with its Karush-Kuhn-Tucker (KKT) conditions \cite{dempe2002foundations}. 
The KKT conditions make the lower problem a mixed-integer linear problem. 
Problem (\ref{equ:single-level-general}) shows the single level equivalent of (\ref{equ:bilevel-general-linear}):

  \begin{subequations}  \label{equ:single-level-general}
    \begin{align}
      \min_{\b x , \b y , \b z, \b \lambda, \overline{\b \mu}, \underline{\b \mu}}  
        & \ f(\b x, \b y, \b z)
        + \frac{a}{b} \sumder \sumt \lambda_{j,t} \ y^{\text{EXP}}_{j,t}  \label{equ:single-level-general-upper-obj}
        \\
      \text{s.t.} & \  
          g(\b x, \b y) \le 0
        \\
          & \
          \b c + \b b^T \b x + \b V^\T \b \lambda + \overline{\b \mu} - \underline{\b \mu} = \b 0 \label{equ:18c}
        \\
          &  \
          \underline{\b{y}} \le \b{y} \le \overline{\b{y}}
        \\
          &  \
          \b{V} \b{y} = \b{w}
        \\
          &  \
          \overline{\b \mu} \perp (\b y - \overline{\b{y}})  \label{equ:comp-con-1}
        \\
          &  \
          \underline{\b \mu} \perp (\underline{\b{y}} - \b y )  \label{equ:comp-con-2}
    \end{align}
  \end{subequations}
  Note that the complementary constraints (\ref{equ:comp-con-1}) and (\ref{equ:comp-con-2}) can be handled with integer variables or special order sets \cite{beale1976global}. Also, the entries of the vector $\b b$ in (\ref{equ:18c}) are zero except for the entries that correspond with the price signal $ x_{j,t}^{\lambda}$. The non-zero values of $\b b$ are set to $\text{pwf}_{\text{LL}}$ in the examples.

Problem \ref{equ:single-level-general} is a mixed integer \textit{non-linear problem}. Using the result from \cite{laws2022linearizing} the product of $\b \lambda$ and $\b y$ in (\ref{equ:single-level-general-upper-obj}) can be linearized as shown in (\ref{equ:linerization}).
Also, if the upper level cost function $f(\b x, \b y)$ and constraints $g(\b x, \b y)$ are linear then the single level problem is mixed-integer linear. However, there are no requirements on the form of the upper level problem within the framework. For example, the constraints in the upper level can represent any power flow equations and integer decisions.

\section{Use-case Examples}
To demonstrate the value of the proposed framework we start with a system that has expected transformer and line over loads. 
The system planner or DSO, as the first player in the upper level, seeks to find the lowest cost solution to the expected over loads by choosing from:
\begin{itemize}
  \item line and transformer upgrades,
  \item battery energy storage systems (BESS),
  \item and/or purchasing DER energy at a time and space varying price of its choosing.
\end{itemize}
The DSO makes these choices with knowledge of how the lower level, i.e. potential DER investors, will react to the price signal. The DER investors, as the second players in the lower level, choose from:
\begin{itemize}
  \item purchasing energy at a known fixed price to meet demand,
  \item and/or investing in DER to meet load as well as (possibly) receive compensation for exported energy.
\end{itemize}

We presume that the DER investors' required rate of return (RoR) is known. The model is formulated such that DER will only be purchased if the RoR can be reached.
The DSO decisions are also subject to power flow constraints, such as line flow and transformer loading limits. 

For reference we compare three scenarios in each use-case:
\begin{enumerate}
  \item The base case with component overloads and the traditional upgrade costs.
  \item The minimum cost solution considering only utility owned BESS (no DER value).
  \item The minimum cost solution considering BESS and valuing DER as a non-wires alternative.
\end{enumerate}

\begin{figure}[!ht]
  \centering
  \includegraphics[width=3.5in]{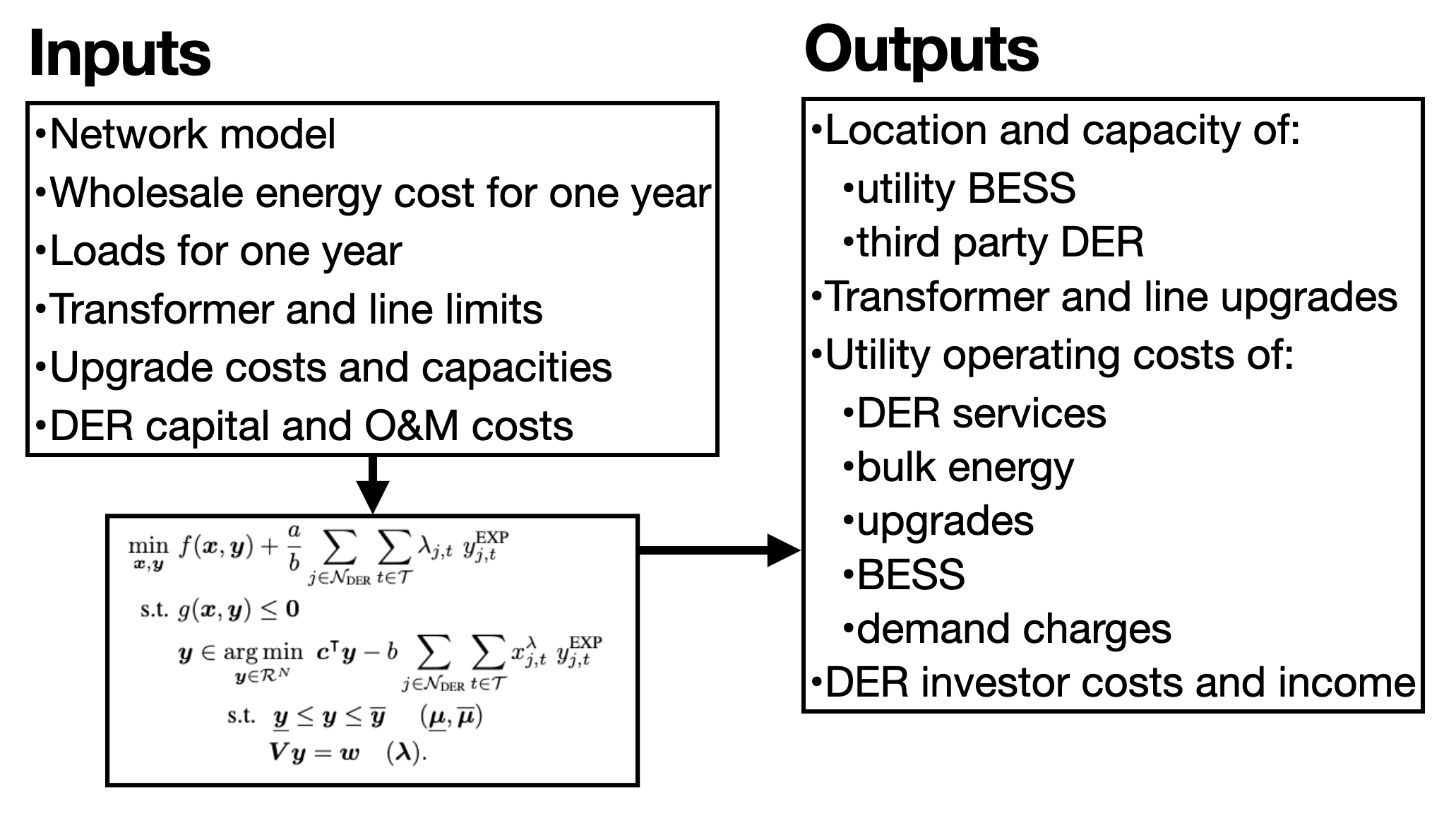}
  \caption{Summary of the use-case example inputs and outputs used to demonstrate the method for valuing DER for non-wires alternatives.}
  \label{fig:ins-outs}
\end{figure}

The input data and the results are summarized in Figure \ref{fig:ins-outs}. 
Note that we assume that the DER are PV systems by using a time-varying production factor from \cite{pvwatts}. 
We presume to have expected hourly load profiles over a year for each load bus in the network\footnote{Random combinations of profiles from \cite{doecrb} were scaled to match the test system.}.
In the base case the utility must pay for new lines and transformers due to expected over loads.
The examples are built upon the IEEE 13 bus distribution test system. Transformer upgrade costs are estimated using values from \cite{nrel2019costs} and line costs were obtained from a large municipal distribution utility.

\begin{figure}[!ht]
  \centering
  \includegraphics[width=3.5in]{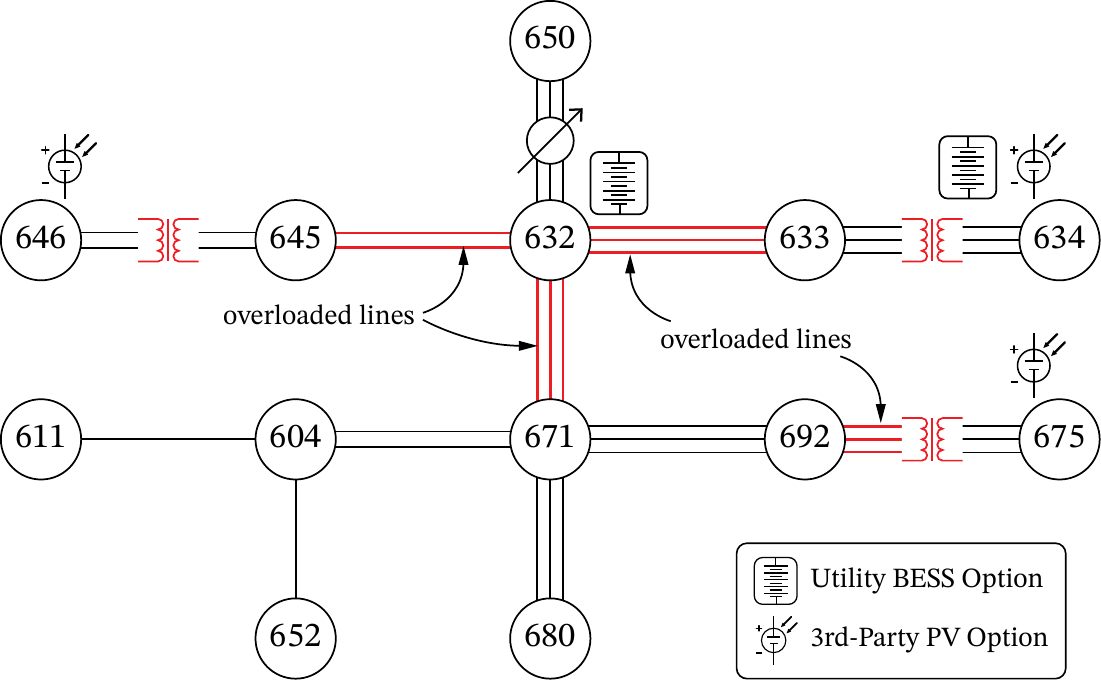}
  \caption{Overview of the IEEE 13 Bus Test System showing the DER and BESS options as well as the overloaded lines and transformers. Secondary transformers at buses 634, 646, and 675 have peak loads of 143\%, 111\%, and 167\% as percent of ratings respectively. Overloaded lines are all assumed to have 110\% overloads compared to their capacity ratings. Graphic by Jeffrey M. Phillips.}
  \label{fig:ntwk}
\end{figure}

Figure \ref{fig:ntwk} gives an overview of the overloaded network components, where the DSO is considering installing BESS, and where it is possible for the DER investors to install PV systems.

\subsection{System Planner Problem}
The upper level problem in the bilevel framework represents the system planner's perspective. 
The upper level cost function $f(\b x, \b y)$ is the sum of battery capital costs, the cost of transformer and line upgrades, the cost of bulk energy purchased, and the peak demand charges:
\begin{equation}  \label{equ:ul-cost}
  \begin{aligned}
    f(\b x, \b y) =& \
          \sum_{j \in \mathcal{N}^{\text{BESS}}, \fap} \left(
              c^{\text{BkW}}  x^{\text{BkW}}_{j,\phi}
          +   c^{\text{BkWh}} x^{\text{BkWh}}_{j,\phi}
          \right)
          \\
          &
          + \sum_{j \in \mathcal{N}^{\text{TRF}} }  c^{\text{TRF}}_{j}  z^{\text{TRF}}_{j}
          + \sum_{j \in \mathcal{N}^{\text{LINE}} } c^{\text{LINE}}_{j} z^{\text{LINE}}_{j}
          \\
          &
          + \text{pwf}_{\text{UL}} \sumt \left( c^{\text{LMP}}_t \sum_{\phi \in \Phi_0}  P_{0,\phi,t}^+ \right)
          \\ 
          & 
          + \text{pwf}_{\text{UL}} \sum_{s \in (1,\dots,|\mathcal{S}|)} c^{\text{DEM}}_s  P_s^{\text{MAX}} 
  \end{aligned}
\end{equation}
The cost coefficients in (\ref{equ:ul-cost}) are defined in Table \ref{tab:ul-cost} and the variables in (\ref{equ:ul-cost}) are defined in Table \ref{tab:ul-vars}.
Note that the net power at the feeder head $P_{0,\phi,t}$ decisions implicitly include lower level decisions via the power flow constraints shown below.
Also, we assume that the DSO is not compensated for exported power by including only the non-negative power $P_{0,\phi,t}^+$ (defined in (\ref{equ:non-neg-P0})) at the feeder head in the objective function.
\begin{subequations} \label{equ:non-neg-P0}
  \begin{align}
  P_{0,\phi,t}^+ \ge 0
  \\
  P_{0,\phi,t}^+ \ge P_{0,\phi,t}, \forall \phi \in \Phi, \fat
  \end{align}
\end{subequations}
DSO are not typically compensated for exported power and can even be restricted to not export at all by contractual agreements or regulations. 

The upper level present worth factor $\text{pwf}_{\text{UL}}$ accounts for annual recurring costs, and is defined as:
\begin{equation}
  a = \text{pwf}_{\text{UL}} = \sum_{y \in 1..N_{\text{years}}} \left(
     \frac{(1+r_e)(1+r_c)}{1+r_{\text{WACC}}}
  \right)^y
\end{equation}
where $r_e$ is the annual cost of electricity growth rate, $r_c$ is the annual energy consumption growth rate, and $r_{\text{WACC}}$ is the planner's weighted average cost of capital rate.

The system planner also has the option to purchase energy from DER (as described in the Section \ref{sec:method}). For these use-case examples we set the coefficient $a$ in (\ref{equ:3}) equal to the system planner's present worth factor $\text{pwf}_{\text{UL}}$. The additional cost component for the planner is thus:
\begin{equation}
  \text{pwf}_{\text{UL}} \sumder \sumt x_{j,t}^{\lambda} \ y^{\text{EXP}}_{j,t}.
\end{equation}

\begingroup
\begin{table}[th!] 
\caption{Parameter values for the distribution system planner in the use-case example.}
\label{tab:ul-cost}
\renewcommand{\arraystretch}{1.3} 
\begin{tabular}{p{0.1\linewidth}p{0.54\linewidth}p{0.2\linewidth}}
  name & description& value \\
  \hline \hline
    $c^{\text{BkW}}$  & cost of BESS inverter [\$/kW]  & 300
  \\  [0.7ex]
    $c^{\text{BkWh}}$  & cost of BESS capacity [\$/kWh]  & 250
  \\  [0.7ex]
    $c^{\text{TRF}}_{j}$  & cost of a new transformer [\$] & 150,000
  \\  [0.7ex]
    $c^{\text{LINE}}_{j}$  & cost of a new line [\$] & length$\times$200+15,000
\\  [0.7ex]
    $c^{\text{LMP}}_t$  & wholesale cost of energy [\$/kWh]  & \cite{ercotrtm}
  \\  [0.7ex]
    $c^{\text{DEM}}_s$  & peak demand cost in period $s$ [\$/kW]  & 50
  \\  [0.7ex]
    $\eta$  & BESS efficiency [fraction] & 0.96
  \\  [0.7ex]
    $\Delta R^{\text{TRF}}_j$  & increase in transformer rating [kW] & varies
  \\ [0.7ex]
    $\overline{R}^{\text{TRF}}_j$ & existing transformer rating [kW] & varies
  \\  [0.7ex]
    $\Delta R^{\text{LINE}}_j$  & increase in line rating [kW] & varies
  \\  [0.7ex]
    $\overline{R}^{\text{LINE}}_j$ & existing line rating [kW] & varies
  \\  [0.7ex]
    $r_e$  & energy cost growth rate [year$^{-1}$]  & 0.03
  \\  [0.7ex]
    $r_e$  & energy consumption growth rate [year$^{-1}$]  & 0.03
  \\  [0.7ex]
    $r_{\text{WACC}}$  & weighted avg. cost of capital [year$^{-1}$]  & 0.10
\\ \hline \\
\end{tabular}
\end{table}
\endgroup

\begingroup
\begin{table}[!ht] 
  \centering
\caption{System planner decision variables for the use-case example with descriptions and units. The planner can invest in BESS at different locations, dispatch the BESS to minimize costs, purchase DER energy, and/or upgrade system components.}
\label{tab:ul-vars}
\renewcommand{\arraystretch}{1.3} 
\begin{tabular}{p{0.5in}p{1.8in}}
  name & description \\
  \hline \hline
    $x^{\text{BkW}}_{j,\phi}$ & BESS inverter capacity [kW]
  \\  [0.7ex]
    $x^{\text{BkWh}}_{j,\phi}$ & BESS energy capacity [kWh]
  \\  [0.7ex]
    $x^{\text{B}^+}_{j,\phi,t}$ & BESS charge rate [kW]
  \\  [0.7ex]
    $x^{\text{B}^-}_{j,\phi,t}$ & BESS discharge rate [kW]
  \\  [0.7ex]
    $x^{\text{SOC}}_{j,\phi,t}$ & BESS state-of-charge [kWh]
  \\  [0.7ex]
    $x_{j,t}^{\lambda}$  & Price of DER exported energy [\$/kWh]
  \\  [0.7ex]
    $z^{\text{TRF}}_{j}$ & binary for transformer upgrade [0/1]
  \\  [0.7ex]
    $z^{\text{LINE}}_{j}$ & binary for transformer upgrade [0/1]
  \\ \hline
\end{tabular}
\end{table}
\endgroup

The system planner considers purchasing BESS systems, which can preclude or reduce the size of system upgrades, lower the cost of bulk energy purchases via energy arbitrage, as well as reduce peak demand charges.
The BESS operational constraints are shown in (\ref{equ:bess-ops}).
\begin{subequations} \label{equ:bess-ops}
  \begin{equation}
  \begin{split}
    x^{\text{SOC}}_{j,\phi,t} = x^{\text{SOC}}_{j,\phi,t-1} +  x^{\text{B}^+}_{j,\phi,t} \eta - x^{\text{B}^-}_{j,\phi,t} / \eta 
    \\ \forall j \in \mathcal{N}_{\text{BESS}},\fap, \forall t \in \mathcal{T} \label{con:firstBESS}
  \end{split}
  \end{equation}
  \\[-30pt]
  \begin{align}
    x^{\text{BkW}}_{j,\phi} \ge x^{\text{B}^+}_{j,\phi,t} + x^{\text{B}^-}_{j,\phi,t},
    \ \forall j \in \mathcal{N}_{\text{BESS}},\fap, \forall t \in \mathcal{T} \label{equ:bess2}
   \\[5pt]
   x^{\text{BkWh}}_{j,\phi} \ge x^{\text{SOC}}_{j,\phi,t},
    \ \forall j \in \mathcal{N}_{\text{BESS}},\fap, \forall t \in \mathcal{T}  \label{equ:bess3}
  \\[5pt]
    x^{\text{SOC}}_{j,\phi,0} = 0.5 x^{\text{BkWh}}_{j,\phi},
    \ \forall j \in \mathcal{N}_{\text{BESS}},\fap  \label{equ:bess4}
  \\[5pt]
    x^{\text{SOC}}_{j,\phi,T} = 0.5 x^{\text{BkWh}}_{j,\phi},
    \ \forall j \in \mathcal{N}_{\text{BESS}},\fap  \label{equ:bess5}
  \\[5pt]
    x^{\text{SOC}}_{j,\phi,t} \geq 0, x^{\text{BkWh}}_{j,\phi} \geq 0,
    \ \forall j \in \mathcal{N}_{\text{BESS}},\fap,\fat \label{con:lastBESS}
  \end{align}
\end{subequations}
In words, the BESS constraints are:
\begin{itemize}
  \item (\ref{con:firstBESS}) defines the time evolution of the battery state of charge;
  \item (\ref{equ:bess2}) says that the sum of the battery power decisions can be at most the inverter rating;
  \item (\ref{equ:bess3}) says that the battery state of charge is at most the battery energy rating;
  \item (\ref{equ:bess4}) says that the initial state of charge is half of the energy rating;
  \item (\ref{equ:bess5}) says that the final state of charge is half of the energy rating; and
  \item (\ref{con:lastBESS}) says that the state of charge and energy rating decisions are non-negative.
\end{itemize}

The system planner has binary decisions for transformer and line upgrades (see Table \ref{tab:ul-vars}).
If a component is upgraded then its operational limits are expanded by the difference in capacity between the original component and the new component. This fact is reflected in Equation (\ref{equ:trfx-limits}) for transformers and Equation (\ref{equ:line-limits}) for lines.
\begin{equation} \label{equ:trfx-limits}
  \begin{split}
  -\overline{R}^{\text{TRF}}_j - z^{\text{TRF}}_{j} \Delta R^{\text{TRF}}_j
   \leq P_{j,\phi,t} \leq 
    \overline{R}^{\text{TRF}}_j + z^{\text{TRF}}_{j} \Delta R^{\text{TRF}}_j,
  \\ \forall j \in \mathcal{J}^{\text{TRF}}, \fap, \fat
  \end{split}
\end{equation}
\begin{equation} \label{equ:line-limits}
  \begin{split}
  -\overline{R}^{\text{LINE}}_j - z^{\text{LINE}}_{j} \Delta R^{\text{LINE}}_j
   \leq P_{ij,\phi,t} \leq 
    \overline{R}^{\text{LINE}}_j + z^{\text{LINE}}_{j} \Delta R^{\text{LINE}}_j,
  \\ \forall j \in \mathcal{J}^{\text{LINE}}, \fap, \fat
\end{split}
\end{equation}

The planner's constraint set also includes a three phase, unbalanced LinDistFlow model \cite{arnold2016optimal}.\footnote{As part of this work an open-source Julia implementation of LinDistFlow was created \cite{lindistflow}.}  The power flow constraints are shown in (\ref{equ:powerflow-cons}).
\begin{subequations} \label{equ:powerflow-cons}
  \begin{equation} \label{equ:pf-Pbalance} \begin{split}
  P_{ij,\phi,t} + P_{j,\phi,t} &- \sum_{k:j\rightarrow k} P_{jk,\phi,t} = 0,
  \\  & \qquad \forall j \in \mathcal{N}, \fap, \fat
  \end{split} \end{equation}
  \\[-25.5pt]
\begin{equation}\label{equ:pf-Qbalance} \begin{split}
  Q_{ij,\phi,t} + Q_{j,\phi,t} &- \sum_{k:j\rightarrow k} Q_{jk,\phi,t} = 0,
  \\  & \qquad \forall j \in \mathcal{N}, \fap, \fat
\end{split} \end{equation}
\\[-25.5pt]
\begin{equation} \begin{split}
    \b v_{j,t} = \b v_{i,t} + \b M^P_{ij} \b P_{ij,t} + \b M^Q_{ij} \b Q_{ij,t},
    \\ \forall (i,j) \in \mathcal{E},\ \forall t \in \mathcal{T}  \label{equ:pf-v}
\end{split} \end{equation}
\\[-25.3pt]
  \begin{align}
  P_{0,\phi,t} = P_{01,\phi,t} 
  \ \forall \phi \in \Phi_0, \fat \label{equ:pf-P0}
  \\ 
  Q_{0,\phi,t} = Q_{01,\phi,t},
  \ \forall \phi \in \Phi_0, \fat \label{equ:pf-Q0}
  \\
  \underline{P}_{j,\phi} \leq P_{j,\phi,t} \leq \overline{P}_{j,\phi}
  \ \forall j \in \mathcal{N}, \fap, \fat \label{equ:pf-Pbounds}
  \\
  \underline{Q}_{j,\phi} \leq Q_{j,\phi,t} \leq \overline{Q}_{j,\phi}
  \ \forall j \in \mathcal{N}, \fap, \fat
  \\
  \underline{P}_{ij,\phi} \leq P_{ij,\phi,t} \leq \overline{P}_{ij,\phi}
  \ \forall (i,j) \in \mathcal{E}, \fap, \fat
  \\
  \underline{Q}_{ij,\phi} \leq Q_{ij,\phi,t} \leq \overline{Q}_{ij,\phi}
  \ \forall (i,j) \in \mathcal{E}, \fap, \fat
  \\
  \underline{v} \leq v_{j,\phi,t} \leq \overline{v},
  \ \forall j \in \mathcal{N}, \ \fap, \ \fat  \label{equ:pf-vbounds}
  \end{align}
\end{subequations}
The real and reactive power balances are shown in (\ref{equ:pf-Pbalance}) and (\ref{equ:pf-Qbalance}) respectively.
Equation (\ref{equ:pf-v}) defines the vector of voltage magnitudes squared for each phase, 
where $\b M^P_{ij}$ and $\b M^Q_{ij}$ are 3$\times$3 matrices of line resistances and reactances (see equations (20) and (21) in \cite{arnold2016optimal}), the vector $\b v_{j,t}=\left[ v_{j,1,t},  v_{j,2,t},  v_{j,3,t}\right]^\T$ collects the phase voltages (squared), and similarly the vectors $\b P_{ij,t}=\left[ P_{ij,1,t}, P_{ij,2,t}, P_{ij,3,t}\right]^\T$ and $\b Q_{ij,t}=\left[ Q_{ij,1,t}, Q_{ij,2,t}, Q_{ij,3,t}\right]^\T$ collect the phase line flows.
Equations (\ref{equ:pf-P0}) and (\ref{equ:pf-Q0}) state that the net power at the feeder head is equal to power transferred along the lines from node zero to one. 
And the remaining constraints (\ref{equ:pf-Pbounds}) -- (\ref{equ:pf-vbounds}) define upper and lower bounds.

The peak demand in each demand period $P_s^{\text{MAX}}$ is defined as the highest demand at the substation in a demand period:
\begin{equation}
  P_s^{\text{MAX}} \geq \sum_{\phi \in \Phi_0} P_{0,\phi,t} \ \forall t \in s, \forall s \in \mathcal{S}
\end{equation}

Lastly, the upper level includes a structural constraint that prevents simultaneous import and export of energy:
\begin{equation}
  y^{\text{IMP}}_{j,\phi,t} \perp y^{\text{EXP}}_{j,\phi,t},
  \ \forall j \in \mathcal{N}_{\text{DER}}, \fap, \fat 
\end{equation}

\subsection{DER Investor Problem}
The potential DER investors have known demands and a time varying cost of energy. Note that if DER costs are low enough (or energy costs high enough) it is possible that DER capacity is installed even in the absence of the price signal. 

The lower level costs are:
\begin{equation}
  \sum_{j \in \mathcal{N}_{\text{DER}}} \left(
    c_j^{\text{kW}}  y_{j}^{\text{kW}} 
  + \text{pwf}_{\text{LL}} \left[  c^{\text{OM}} y_{j}^{\text{kW}} + \sumt c^{\text{IMP}}_{j,t} y^{\text{IMP}}_{j,t}  \right]
  \right),
\end{equation}
which in words is the sum over all potential DER investor nodes $\mathcal{N}_{\text{DER}}$ of:
\begin{itemize}
  \item the capital cost of DER $c_j^{\text{kW}}  y_{j}^{\text{kW}}$,
  \item the operations and maintenance cost of DER $c^{\text{OM}} y_{j}^{\text{kW}}$,
  \item and the cost of energy $c^{\text{IMP}}_{j,t} y^{\text{IMP}}_{j,t}$.
\end{itemize}
The DER investor parameters and variables are summarized in Tables (\ref{tab:ll-cost}) and (\ref{tab:ll-vars}) respectively.

The lower level benefits (negative costs) are the sum of the price signal $x_{j,t}^{\lambda}$ from the upper level and the exported energy $y^{\text{EXP}}_{j,t}$ over time and space:
\begin{equation} \label{equ:ll-income}
  - \text{pwf}_{\text{LL}} \sumder \sumt x_{j,t}^{\lambda} \ y^{\text{EXP}}_{j,t}.
\end{equation}
Note that the bilinear product of upper and lower level decisions in (\ref{equ:ll-income}) are linearized when the lower level problem is replaced with its KKT conditions (see (\ref{equ:single-level-general})).

The lower level present worth factor $\text{pwf}_{\text{LL}}$, defined in (\ref{equ:pwfll}), accounts for the:
\begin{itemize}
  \item years of the analysis period $N_{\text{years}}$,
  \item the RoR,
  \item the annual cost of energy growth rate $r_e$,
  \item and the annual energy consumption growth rate $r_c$.
\end{itemize}

\begin{equation} \label{equ:pwfll}
  b = \text{pwf}_{\text{LL}} =
     \sum_{y \in 1..N_{\text{years}}} \left(
       \frac{(1+r_e)(1+r_c)}{1+\text{RoR}}
     \right)^y
\end{equation}

\begingroup
\begin{table}[!ht] 
  \centering
\caption{DER investor baseline parameters for the use-case example.}
\label{tab:ll-cost}
\renewcommand{\arraystretch}{1.3} 
\begin{tabular}{p{0.5in}p{2.1in}p{0.3in}}
  name & description& value \\
  \hline \hline
    $c^{\text{kW}}_j$  & cost of DER at node $j$ [\$/kW]  & 1,600
  \\  [0.7ex]
    $c^{\text{OM}}$  & annual cost of operations and maintenance [\$/(kW-year)]  & 17
  \\  [0.7ex]
    $c^{\text{IMP}}_{j,t}$  & cost of energy imported [\$/kWh]  & 0.15
  \\ [0.7ex]
    $d_{j,t}$  & uncontrollable demand [kW]  & \cite{crb}
  \\ [0.7ex]
    $f_{j,t}^{\text{prod}}$  & production factor, 0--1 & \cite{pvwatts}
  \\  [0.7ex]
    $r_e$  & energy cost growth rate [year$^{-1}$]  & 0.03
  \\  [0.7ex]
    $r_e$  & energy consumption growth rate [year$^{-1}$]  & 0.03
  \\  [0.7ex]
    $ROR$  & required rate of return [year$^{-1}$]  & 0.15
\\ \hline \\
\end{tabular}
\end{table}
\endgroup

\begingroup
\begin{table}[!ht] 
  \centering
\caption{DER investor variables for the use-case example. The investor chooses how much capacity to install as well as how to dispatch the system.}
\label{tab:ll-vars}
\renewcommand{\arraystretch}{1.3} 
\begin{tabular}{p{0.5in}p{1.7in}}
  name & description \\
  \hline \hline
    $y_{j}^{\text{kW}}$  & capacity of DER at node $j$ [kW]
  \\  [0.7ex]
    $y^{\text{IMP}}_{j,t}$  & energy imported [kWh]
  \\  [0.7ex]
    $y^{\text{EXP}}_{j,t}$  & energy exported [kWh]
  \\  [0.7ex]
    $y^{\text{DER}}_{j,t}$  & energy produced [kWh]
\\ \hline \\
\end{tabular}
\end{table}
\endgroup

The DER investor load balance defines the net power injection for each node and time step:
\begin{equation}
  d_{j,t} - y^{\text{DER}}_{j,t} = y^{\text{IMP}}_{j,t}  - y^{\text{EXP}}_{j,t}
  \ (\lambda_{j,t}),
  \ \forall j \in \mathcal{N}_{\text{DER}}, \fat 
\end{equation}

Lastly, the real power production of any DER is limited by the purchased capacity and a time-varying production factor (between zero and one):
\begin{equation}
 y^{\text{DER}}_{j,t} \le y^{\text{kW}}_{j} f_{j,t}^{\text{prod}}, 
 \ \forall j \in \mathcal{N}_{\text{DER}}, \fat 
\end{equation}

\section{Results}
\begin{figure}[!ht]
  \centering
  \includegraphics[width=3.5in]{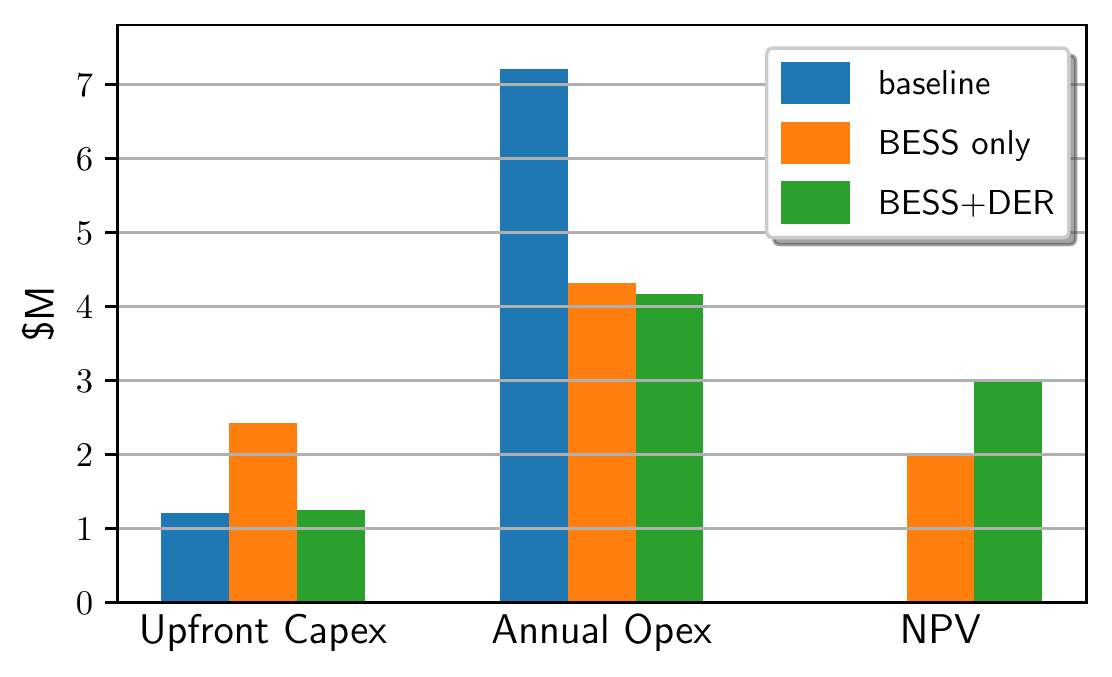}
  \caption{Case study summary results. See Table \ref{tab:results_summary} for a break down of the upfront capital costs (capex) and the annual operating costs (opex). The net present value (NPV) is by definition zero in the baseline case.}
  \label{fig:results_summary}
\end{figure}

Figure \ref{fig:results_summary} provides a high level summary of the results.
In the baseline case the system planner must upgrade the overloaded components and cannot invest in BESS nor purchase DER energy. 
The total lifecycle cost (LCC) for the planner is \$8.41M over 20 years of operations, including the capital costs of upgrades.
A break down of the planner's baseline costs are summarized in Table \ref{tab:results_summary}. 

\begin{figure}[!ht]
\centering
\includegraphics[width=3.5in]{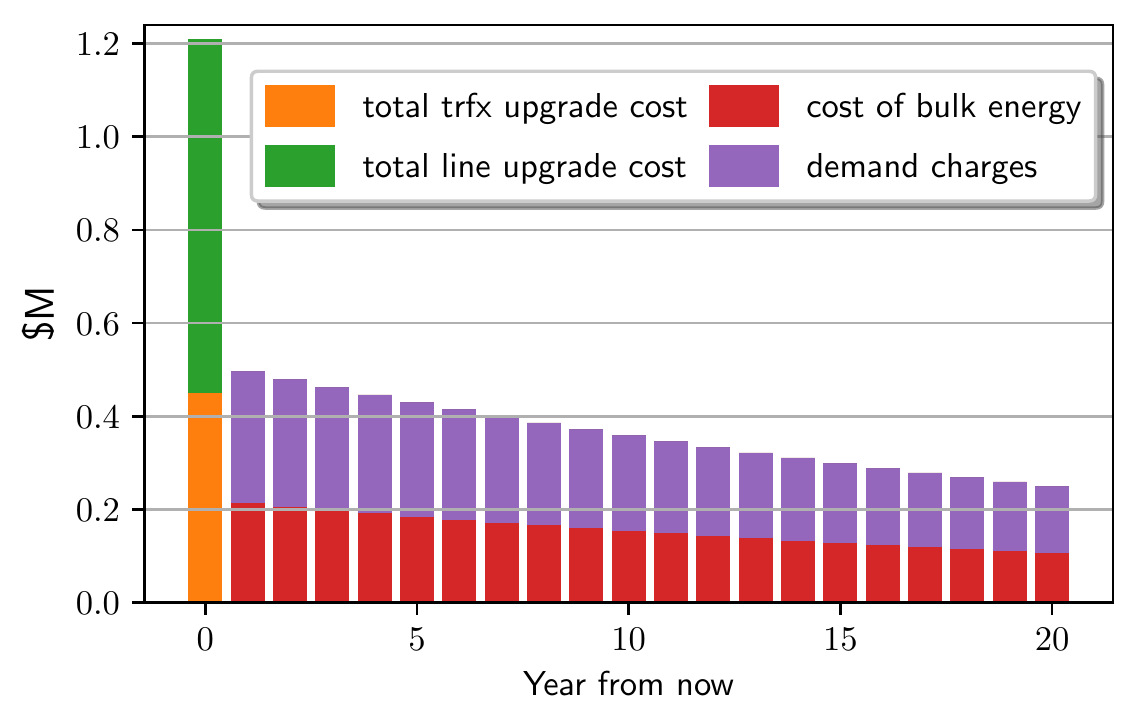}
\caption{Upfront capital costs (in year zero) and annual, discounted operating costs for the DSO in the base case with traditional upgrades.}
\label{fig:cashflow_base_cannot_sell}
\end{figure}

Figure \ref{fig:cashflow_base_cannot_sell} shows the annual costs in the base scenario. Note that the first bar represents the upfront capital costs, which in the base case include the traditional upgrade costs for the lines and transformers with expected over loads. The annual operating costs include the cost of bulk energy as well as the coincident peak demand charges.

\begingroup
\begin{table}[th!] 
\caption{A summary of the use-case example results for the baseline, BESS only, and BESS with DER valued as non-wires alternatives. All dollar values are in millions. (Abbreviations: ``LCC" = lifecycle cost, ``Trfx" = transformer, ``capex" = capital cost.)}
\label{tab:results_summary}
\renewcommand{\arraystretch}{1.3} 
\begin{tabular}{p{0.24\linewidth}p{0.17\linewidth}p{0.17\linewidth}p{0.2\linewidth}}
                & Baseline & BESS only & BESS \& DER \\
  \hline \hline
    Total LCC      & \$8.41 & \$6.43 & \$5.42
  \\  \hline
    Net present value  & -- & \$1.98 & \$2.99
  \\  
    Trfx upgrades  & \$0.45 & \$0.30 & \$0.15
  \\  
    Line upgrades  & \$0.76 & \$0.65 & \$0.12
  \\  
    Bulk energy    & \$3.09 & \$1.70 & \$1.17
  \\ 
    Demand charges & \$4.11 & \$2.61 & \$2.25
  \\
    BESS capex     & --      & \$1.47 & \$0.96
  \\  
    DER energy     & --      &  --     & \$0.75
  \\  \hline
    Lines upgraded & 4/4     &  3/4    & 1/4
  \\  
   Trfxs upgraded  & 3/3     &  2/3    & 1/3
\\ \hline 
\end{tabular}
\end{table}
\endgroup

Next, we present the results for the system planner when considering only BESS investments (without purchasing DER energy). 
The DSO can install BESS at nodes 632 and/or 634 (see Figure \ref{fig:ntwk}). 
In this scenario it is cost optimal for the DSO to install a 4.4 hour, 21.8 kW BESS at bus 634 and a 4.8 hour, 961 kW BESS at bus 632.
With these BESS the total life cycle cost is \$6.43M, yielding \$1.98M in savings over the base scenario. 
The addition of the BESS also prevent the need to upgrade one of the lines (632-633) as well as the transformer at bus 634.
\begin{figure}[!ht]
\centering
\includegraphics[width=3.5in]{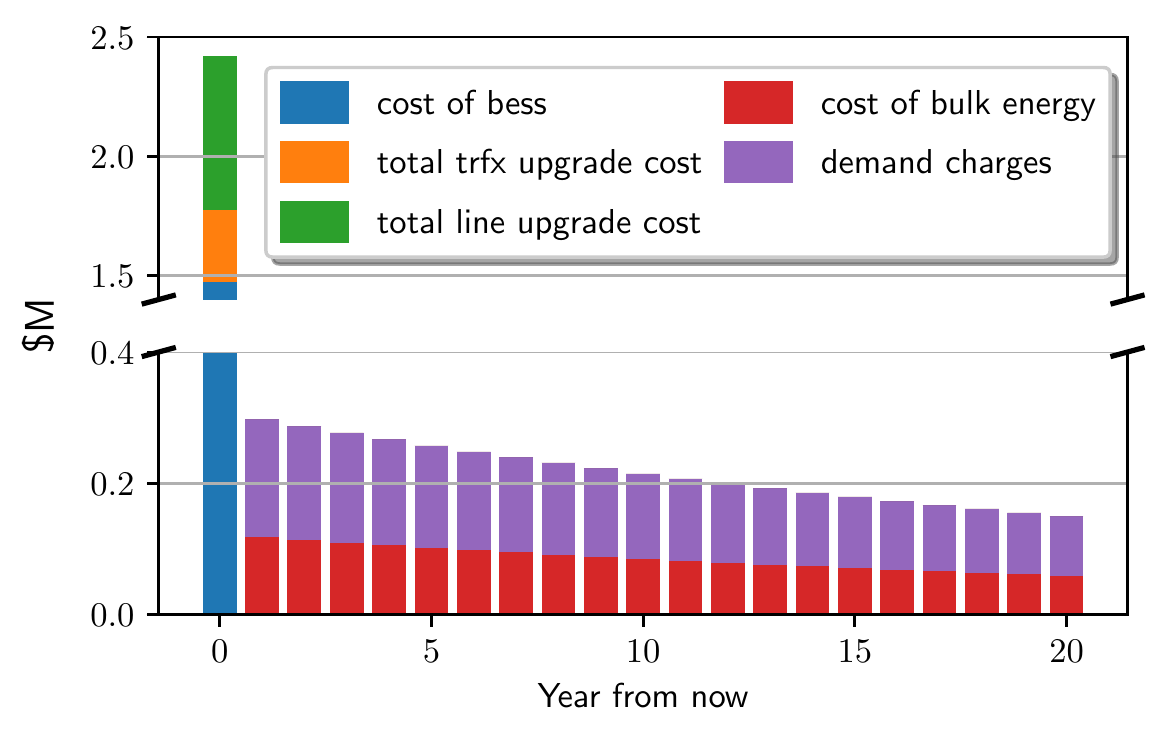}
\caption{Upfront capital costs (in year zero) and annual, discounted operating costs for the DSO considering only BESS (no DER). Note the much higher upfront costs when compared to the base line upfront costs in Figure \ref{fig:cashflow_base_cannot_sell} come with the benefit of lower annual operating costs as compared to the base scenario.}
\label{fig:cashflow_bess_only_cannot_sell}
\end{figure}

Comparing the cash flows with and without BESS in Figures \ref{fig:cashflow_base_cannot_sell} and \ref{fig:cashflow_bess_only_cannot_sell} respectively, yields two key observations: i) the planner's upfront capital costs are approximately doubled with the BESS investment, even though the line and transformer upgrade costs are reduced relative to the base case; and ii) the first year operating costs are reduced by approximately \$0.2M with the optimal BESS.

In the third and final scenario we include the value of DER for non-wires alternatives.
Now, with the option to purchase DER energy the DSO total lifecycle cost is \$5.42M, yielding an additional cost reduction of \$1.01M over the BESS only scenario for a total net present value of \$2.99M. 
The DSO reduces its cost of energy by purchasing DER exported energy at the DER investor's marginal cost, in this case 15 cents/kWh\footnote{We do not use time-varying retail rates to keep the use-case examples relatively easy to understand. However, there is no limit on the form of the retail rates.}. 
Additionally, with the DER contributions the DSO only needs to upgrade one of the three overloaded transformers (at bus 675) and one set of the four overloaded lines (692-675).
Table \ref{tab:results_summary} summarizes the costs when valuing DER.

\begin{figure}[!ht]
\centering
\includegraphics[width=3.4in]{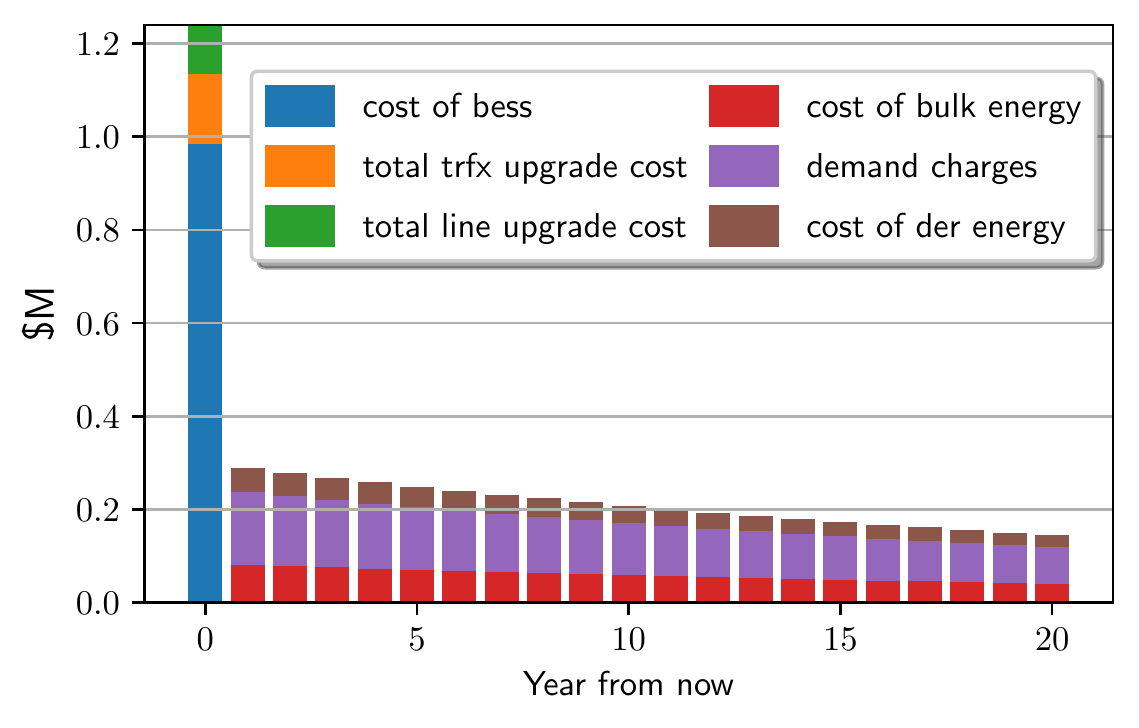}
\caption{Upfront capital costs (in year zero) and annual, discounted operating costs for the DSO considering BESS and DER for non-wires alternatives. The upfront capital costs are comparable to the traditional upgrade costs and much lower than the upfront costs in the BESS only scenario shown in Figure \ref{fig:cashflow_bess_only_cannot_sell}. Also, the annual operating costs are much lower than the baseline scenario shown in Figure \ref{fig:cashflow_base_cannot_sell} and lower than the BESS only scenario, even with the additional cost of purchasing DER energy.}
\label{fig:cashflow_bess_der_cannot_sell}
\end{figure}

Figure \ref{fig:cashflow_bess_der_cannot_sell} shows the annual cash flows for the third scenario with DER valued. Note that the upfront costs are comparable to the base scenario but the annual operating costs are significantly reduced when compared to the base scenario cash flows in Figure \ref{fig:cashflow_base_cannot_sell}.

For reference we also provide the DER investor results with and without the DSO price signal in Table \ref{tab:der_results}. Note that the net present cost values are the same with and without the price signal. In both cases the DER investors obtain their required rate of returns. However, in the case with the price signal the additional costs of the larger DER systems are offset with the income from selling energy as well as the additional energy cost savings.
\begingroup
\begin{table}[!ht] 
\caption{Use-case example results summary for DER investors with and without the price signal from the DSO. All dollar values are in millions. (Abbreviations: ``capex" = capital cost.)}
\label{tab:der_results}
\renewcommand{\arraystretch}{1.3} 
\begin{tabular}{p{0.4\linewidth}p{0.2\linewidth}p{0.2\linewidth}}
                   & no signal & with signal  \\
  \hline \hline
    Net present cost   & \$2.79 & \$2.79
  \\  \hline
    DER Capex  & \$0.43 & \$1.59
  \\  
    Income    & -- & \$0.51
  \\  
    Energy cost savings    & \$0.53 & \$1.30
  \\  
    Bus 634 capacity  & 67 kW & 138 kW
  \\ 
    Bus 646 capacity  & 124 kW & 305 kW
  \\  
    Bus 675 capacity  & 79 kW & 550 kW
\\ \hline 
\end{tabular}
\end{table}
\endgroup

\section{Discussion}
Using the proposed framework for valuing DER as non-wires alternatives, the use-case examples demonstrate the potential for reducing system operating costs in ways that benefit the DSO as well as DER investors. 
Ultimately, it is reasonable to expect that lower system costs will lead to lower costs for all customers.

The use-case examples assumed that DER investors consider PV systems since they are cost-competitive in many areas and are not subject to emission regulations like back-up gas generators. However, back-up gas generators could provide a significant value proposition for both the DSO and DER investors by providing power during evening peak loads (unlike PV systems) as well as serve customer critical loads during outages. The proposed framework for valuing DER can account for gas generators by removing the time varying limit on the DER production and adjusting the cost parameters. 

All scenarios were solved on a 2017 Macbook Pro with two each 2.8 GHz Quad-Core Intel i7 chips, 16 GB of RAM, and Gurobi version 9. 
The bilevel problem solve time was limited to one hour, in which it reached an optimality gap of less than two percent. 
While the long solve time may indicate that the bilevel method may not scale well to large problems,
no efforts were made to make the bilevel problem easier to solve. 
Future work could include appropriately scaling the problem coefficients as well as decomposing the problem into sub-problems and leveraging advanced solution techniques.

A major advantage of the proposed framework is its flexibility.
For example, it is straightforward to account for different DER types in the lower level problem, as long as they can be modeled in a linear fashion.
The upper level problem is not limited to linear equations, as demonstrated in the use-case example with the inclusion of integer decisions for the transformer and line upgrades.
Another example of leveraging the flexibility of the framework would be to replace the non-negative bulk power purchases $P_{0,\phi,t}^+$ with the net power $P_{0,\phi,t}$ in the upper level objective (\ref{equ:ul-cost}). Valuing power exported to the bulk system would increase the value of DER for non-wires alternatives by adding another value stream for the DSO. 
For example, it is possible that DSO could sell excesss energy in a bulk market.

\section{Conclusion}
Valuing DER for non-wires alternatives appropriately is a difficult task. The framework proposed in this work accounts for both the system planner's perspective and the DER investor perspectives. The bilevel optimization framework guarantees that solutions minimize the planner's costs over the chosen horizon as well that the DER investors achieve their required rate of returns.  In light of FERC Order 2222 \cite{ferc2222} and expected growth in load and DER adoption \cite{eia2022aeo} it is becoming more and more important for system planners to work with DER investors to plan efficient distribution power systems.

Using a use-case example we showed how the framework can be leveraged to value DER for non-wires alternatives. Comparing life cycle costs over 20 years for the system planner, the results show that by valuing DER for non-wires alternatives the DSO can avoid upgrading most of the over loaded components as well as achieve a net present value of nearly \$3M relative to the cost of the traditional upgrades. We also compared the solution with DER valued to a scenario with utility owned batteries and no third-party DER value. The results show that the DSO can achieve an additional \$1M in net present value when valuing DER relative to the scenario with utility owned batteries.
  
In future work we intend to leverage the bilevel framework in a transactive control context. 
Transactive control methods that account for the DSO perspective and the DER owner objectives are necessary to appropriately motivate DER to provide services that benefit the entire system.


%




\ifCLASSOPTIONcaptionsoff
  \newpage
\fi



\bibliographystyle{IEEEtran}
\bibliography{planning.bib}

\vspace{-12mm}
%



%

\begin{IEEEbiography}[{\includegraphics[width=1in,height=1.25in,clip,keepaspectratio]{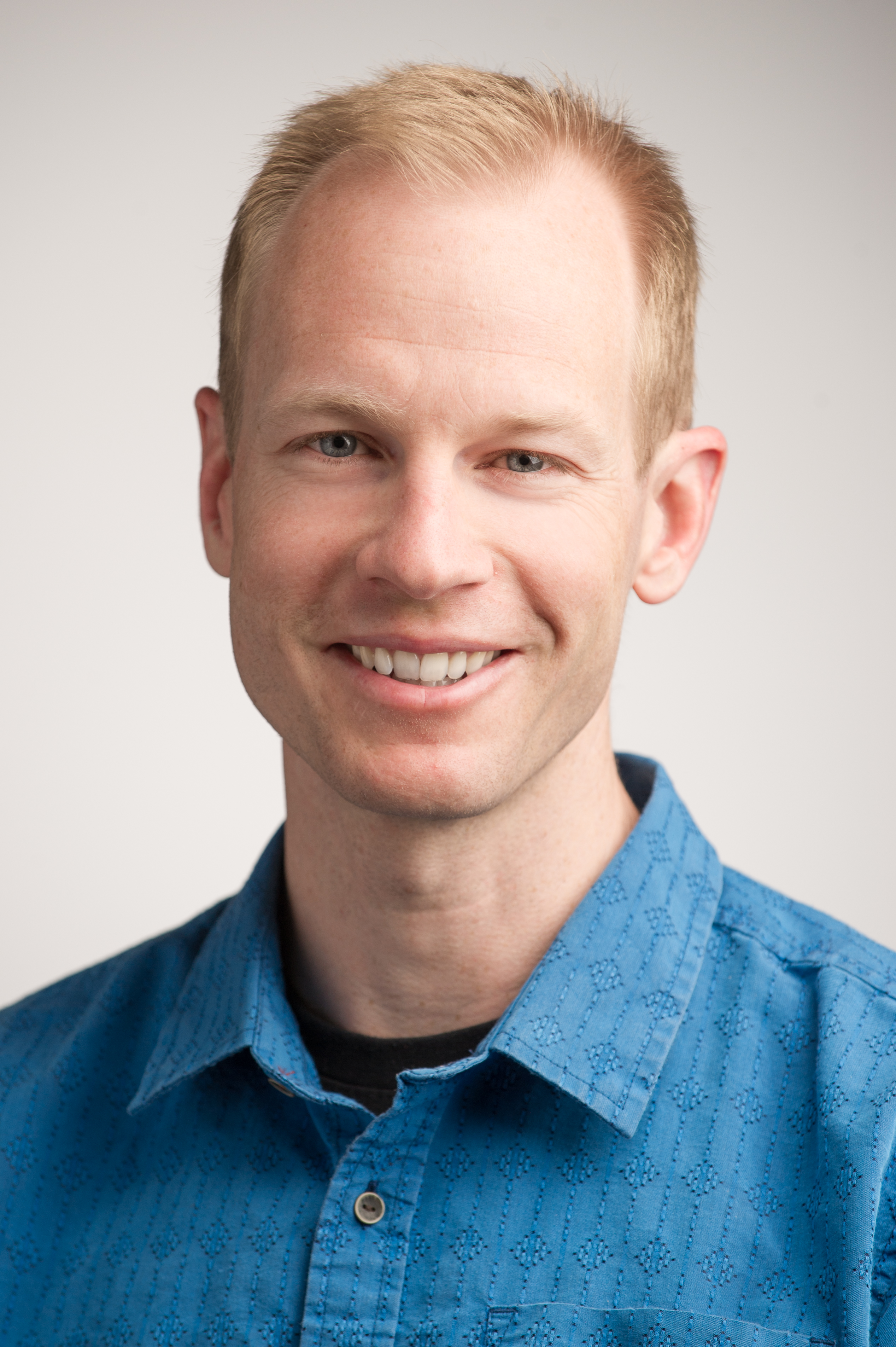}}]{Nicholas D. Laws}
Nicholas D. Laws is a PhD candidate in the Webber Energy Group at The University of Texas at Austin and a Power Systems Optimization Engineer at Camus Energy. His research focuses on the integration of clean distributed energy resources in ways that benefit all stakeholders. He holds a Bachelor of Science in Aerospace Engineering from Boston University and a Master of Science from Dartmouth College, Thayer School of Engineering.
\end{IEEEbiography}

  





\end{document}